\documentclass[journal]{IEEEtran}
\usepackage{graphicx}

\begin{document}
\title{Performance of Irradiated Thin Edgeless N-on-P\\
Planar Pixel Sensors for ATLAS Upgrades}

\author{Marco~Bomben, 
Alvise~Bagolini, Maurizio~Boscardin, Luciano~Bosisio, Giovanni~Calderini, Jacques~Chauveau, Gabriele~Giacomini, Alessandro~La~Rosa, Giovanni~Marchiori and~Nicola~Zorzi% <-this % stops a space
\thanks{Manuscript received November 15, 2013.}% <-this % stops a space
\thanks{M.~Bomben is with Laboratoire de Physique Nucleaire et de Hautes \'Energies (LPNHE), 75252 PARIS CEDEX 05, France (e-mail: marco.bomben@cern.ch).}%
\thanks{A.~Bagolini, M.~Boscardini, G.~Giacomini and N.~Zorzi are with Fondazione Bruno Kessler, Centro per i Materiali e i Microsistemi (FBK-CMM), 38123 Povo di Trento (TN),
Italy.}%
\thanks{L.~Bosisio is with Universit\`a di Trieste, Dipartimento di Fisica and INFN, 34127 Trieste, Italy.}%
\thanks{G.~Calderini is with Laboratoire de Physique Nucleaire et de Hautes \'Energies (LPNHE), 75252 PARIS CEDEX 05, France,  and Dipartimento di Fisica E. Fermi, Universit\`a di Pisa, and INFN Sez. di Pisa, 56127 Pisa, Italy}%
\thanks{J.~Chauveau and G.~Marchiori are with Laboratoire de Physique Nucleaire et de Hautes \'Energies (LPNHE), 75252 PARIS CEDEX 05, France.}%
\thanks{A.~La Rosa is with Section de Physique (DPNC), Universit\`e de Gen\`eve, CH-1211 Gen\`eve 4, Switzerland}%
}

\maketitle
\pagestyle{empty}
\thispagestyle{empty}

\begin{abstract}
In view of the LHC upgrade phases towards the High Luminosity LHC (HL-LHC), the ATLAS
experiment plans to upgrade the Inner Detector with an all-silicon system. Because of its
radiation hardness and cost effectiveness, the n-on-p silicon technology is a promising
candidate for a large area pixel detector. The paper reports on the joint development, by LPNHE and FBK of novel n-on-p edgeless
planar pixel sensors, making use of the active trench concept for the reduction of the dead
area at the periphery of the device. After discussing the sensor technology, a complete
overview of the electrical characterization of several irradiated samples will be discussed. Some comments about detector modules being assembled 
will be made and eventually some plans will be outlined. 
\end{abstract}


\section{Introduction}
\IEEEPARstart{I}{n} the next decade the CERN Large Hadron Collider (LHC) should be upgraded to the so-called High Luminosity  LHC (HL-LHC)~\cite{Rossi:1471000}, capable 
of a luminosity of  $5\times10^{34} {\rm cm}^{-2}{\rm s}^{-1}$, to extend its physics reach. 
By then the ATLAS collaboration will be equipped with a completely new Pixel Detector.  
The innermost layer of the new pixel detector will integrate a fluence of about $10^{16}\, {\rm 1\, MeV\, n_{eq}}/{\rm cm}^2$ for an integrated luminosity of 2500 fb$^{-1}$ 
($\sim$~10 years of operation).
These harsh conditions demand radiation-hard devices and a finely segmented detector to cope with the expected 
high occupancy.
The new pixel sensors will need to have  
  high geometrical acceptance: the future material budget restrictions and  
 tight mechanical constraints require the geometric inefficiency to be less than  2.5\%~\cite{IBLTDR}.
One way to reduce or even eliminate the insensitive region along the device periphery is offered by
 the ``active edge'' technique~\cite{3dKenney}, in which a deep vertical trench is etched along the device periphery throughout the entire wafer thickness, 
thus performing a damage free cut (this requires using a support wafer, to prevent the individual chips from getting loose). 
The trench is  then heavily doped, extending the ohmic back-contact to the lateral sides of the device: the depletion region can then extend to the edge without causing 
a large current increase.
 This is the technology that was chosen  for realizing n-on-p pixel sensors with reduced inactive zone whose features and measurement results are reported in this paper. 
 
 In Section~\ref{sec:prod} the active edge technology chosen for a first production of n-on-p sensors is briefly presented. 
In Section~\ref{sec:irrad} the results from the electrical characterization of  irradiated samples are discussed. 
The preparation of pixel sensors for bump-bonding will be commented in Section~\ref{sec:fei4}.
Eventually, in Section~\ref{sec:concl} some conclusions will be drawn and future plans will be outlined.

\section{The active edge production}\label{sec:prod}

All the details about the sensor production and the active edge technique were already reported in~\cite{bib:nim2012}. Hence  only major features will be 
shortly outlined in this paper. 

\begin{figure}[!t]
\centering
\includegraphics[width=3.5in]{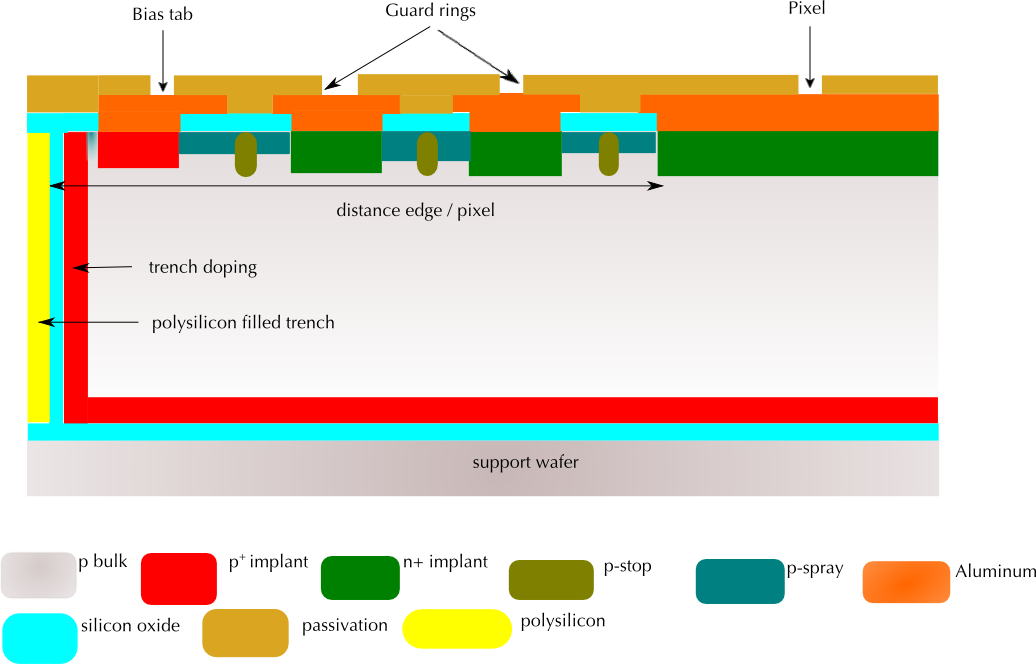}
\caption{\label{fig:design}Schematic section of the pixel sensor. The region close to the sensor's edge is portrayed, including the pixel closest to the edge,
the edge region, including GRs (when present), the bias tab (present only on one edge of the device), the vertical doped trench, and the support wafer.}
\end{figure}

The sensors are fabricated on 100~mm diameter, high resistivity, p-type, Float Zone (FZ), \textless100\textgreater\, oriented, 200~${\rm \mu}$m thick wafers. 
The active edge technology is used, which is a single sided process, featuring a doped trench, 
extending all the way through the wafer thickness, 
and completely surrounding the sensor. For mechanical reasons, a support wafer is  needed.
  Nine FE-I4 readout chip~\cite{FEI4} compatible pixel sensors were put on each wafer; each sensor consists of an array of 336~$\times$~80 pixels, at a pitch  
of 50~${\rm \mu}$m~$\times$~250~${\rm \mu}$m, 
for an overall sensitive area of 16.8~mm~$\times$~20.0~mm. 
The nine FE-I4 sensors differ in the pixel-to-trench distance (100, 200, 300, and 400~${\rm \mu}$m) and in the number of the guard rings (0, 1, 2, 3, 5, and 10)  
surrounding the pixel area. The sensor with 3 GRs and a 200~$\mu$m pixel-to-trench distance features two different GR designs, and 
each of them is repeated twice. A simplified layout of the sensors' design is reported in Figure~\ref{fig:design}.
A list of the different FE-I4 sensor versions is reported in Table~\ref{tab:fei4_devices}.

\begin{table}[!ht]
\caption{\label{tab:fei4_devices}The list of the different FE-I4 compatible-sensor layouts. 
Two different designs are envisaged for the sensor with 3 GRs and 200~$\mu$m pixel-to-trench distance. See text for more details.}
\centering
\begin{tabular}{ccc}
name & \# of GRs & pixel-to-trench distance (${\rm \mu m}$) \\
\hline
S1 &  0 & 100 \\
S2 & 2 & 100 \\ 
S3 & 1 &100\\
S4 & 3 & 200\\ 
S5 & 3 & 200\\
S6 & 3 & 200\\
S7 & 3 & 200\\ 
S8 & 5 & 300\\ 
S9 & 10 & 400 
\end{tabular}
\end{table}

\begin{figure}[!htb]
\begin{center}
\includegraphics[width=3.5in]{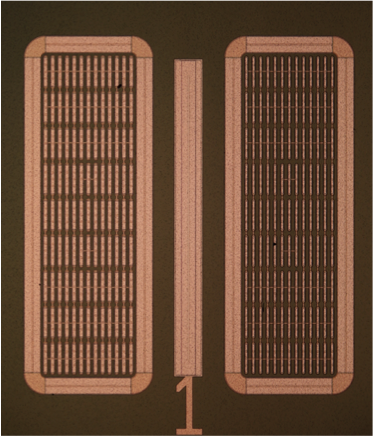}
\includegraphics[width=3.5in]{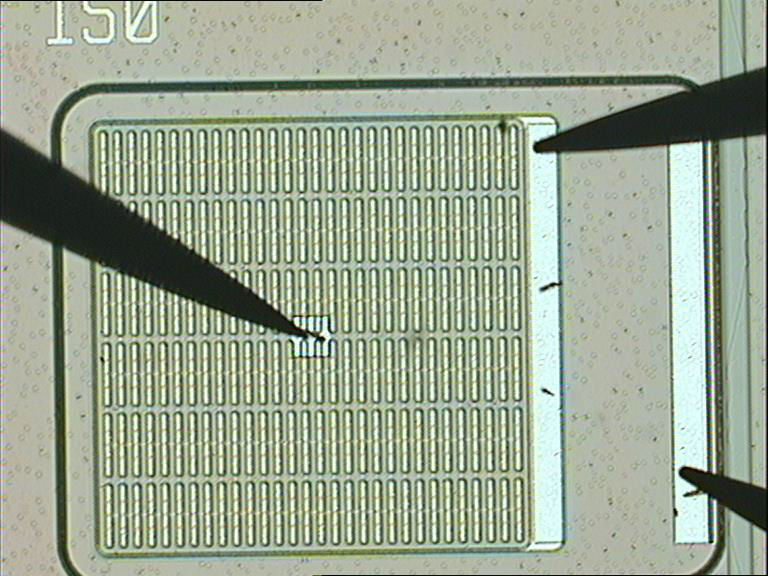}
\caption{\label{fig:pix_cap_struct}Top: test structures consisting of 2 arrays of 9~$\times$~13 FE-I4-like pixel cells each (``inter-pixel structure''); 
the pixels in the left (right) structure have (no) field-plate. 
Bottom: test structures consisting of an array of 6~$\times$~30 FE-I4-like
 pixel cells (``FE-I4 test structure''), where  all the pixels were shorted together.}
\end{center}
\end{figure}

The first part of the measurement program was carried   on squared pad diodes and on structures reported in Figure~\ref{fig:pix_cap_struct}. 
 A test structure consisting of an array 
 of 9~$\times$~13 FE-I4-like pixel cells was used to measure the inter-pixel  and the pixel-to-backside capacitance; the central pixel was isolated with respect to all the other pixels;
 the first 8 neighbours were shorted together, but isolated from all the other remaining (which, again,  were shorted together). These structures are shown on the 
 left in Figure~\ref{fig:pix_cap_struct}, where two versions are present: one with metal field-plate and without. ``inter-pixel structure'' will be used for the sake of 
 brevity in the remaining of the text to refer to this structure. 
 In Figure~\ref{fig:pix_cap_struct}, on the right, an array of 6~$\times$~30 FE-I4-like
 pixel cells is shown; all the pixels were shorted together allowing the measurement of the current voltage characteristics  of the whole array and of the inner GR (if present), 
 and the breakdown (BD) voltage dependence on 
 the number of GRs and on the pixel-to-trench distance. Several combinations of values for the latter parameters are present on the wafer; in 
 Figure~\ref{fig:pix_cap_struct}, on the right, a structure with a 200~$\mu$m pixel-to-trench distance and 2 GRs is shown. ``FE-I4 test structure'' will be used for the sake of 
 brevity in the remaining of the text to refer to this structure. 
 
The results of the production electrical characterization before irradiation have been reported in~\cite{2013arXiv1311.1628B}. In the following Section the 
results after irradiation are presented.

\section{First results from irradiated samples}\label{sec:irrad}

In order to assess the radiation hardness of the production, several samples were irradiated at the Triga reactor~\cite{Triga}. The total integrated fluence $\Phi$  was of  
$2.5 \times 10^{15}\, {\rm (1\, MeV)\, n_{eq}}/{\rm cm}^2$.  Very limited annealing at room temperature was experienced by the samples after the irradiation. All the measurements 
were carried at a temperature t=$(0\pm1)^\circ {\rm C}$. All the sensors were taken from a single wafer, which featured a uniform p-spray implant 
(dose~$\sim$ 3$\times$10$^{12}$/cm$^2$) and p-stops to ensure pixels' isolation.

Bulk properties after irradiation were studied using simple pad diodes with 2 GRs.
Surface properties as well as the BD voltage 
 were studied making use of the test structures reported in Figure~\ref{fig:pix_cap_struct} and discussed before. 

\begin{figure}[!t]
\centering
\includegraphics[width=3.5in]{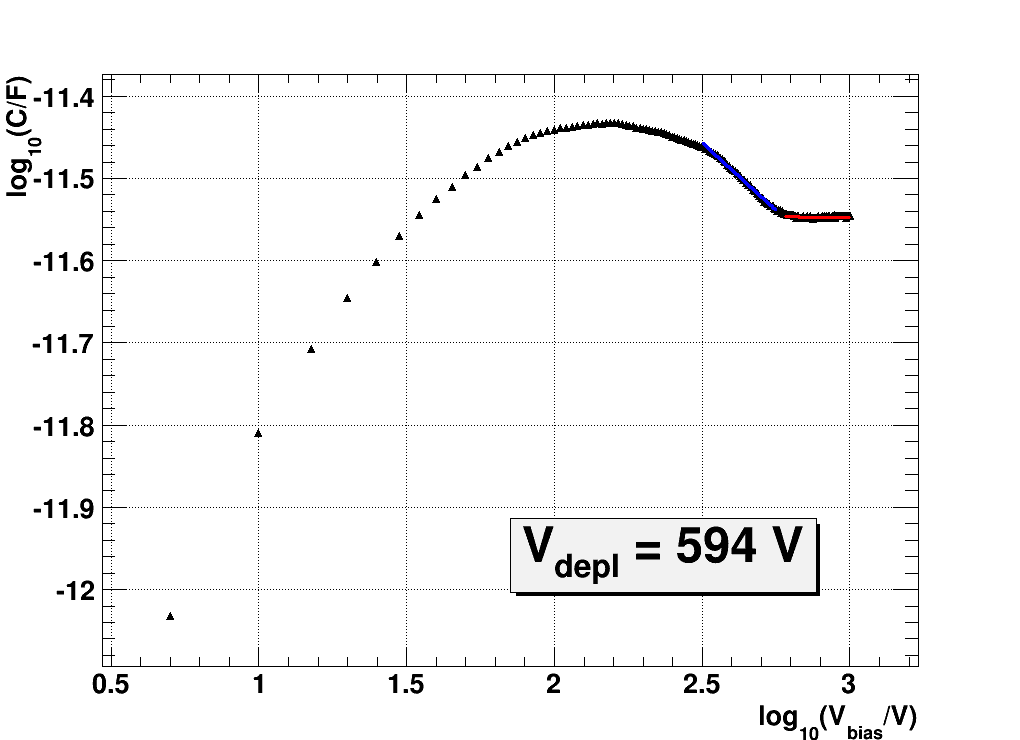}
% where an .eps filename suffix will be assumed under latex, 
% and a .pdf suffix will be assumed for pdflatex; or what has been declared
% via \DeclareGraphicsExtensions.
\caption{Capacitance as a function of bias voltage for the irradiated squared pad diode.  Both GRs and the pad were kept at ground potential while the backside was 
biased via a p$^+$ implant at the detector periphery. The crossing point of 2 straight lines (in red and blue in the graph) was used as an estimate of the depletion voltage.}
\label{fig:irrCV}
\end{figure}

The depletion voltage was estimated by studying the pad-to-back capacitance as a function of the bias voltage; both GRs were kept at the same potential of the pad 
and a 10~kHz frequency was used. Results are reported in Figure~\ref{fig:irrCV}. 
A depletion voltage value of about 600~V was found, in agreement 
with literature~\cite{moll-thesis}.

\begin{figure}[!t]
\centering
\includegraphics[width=3.5in]{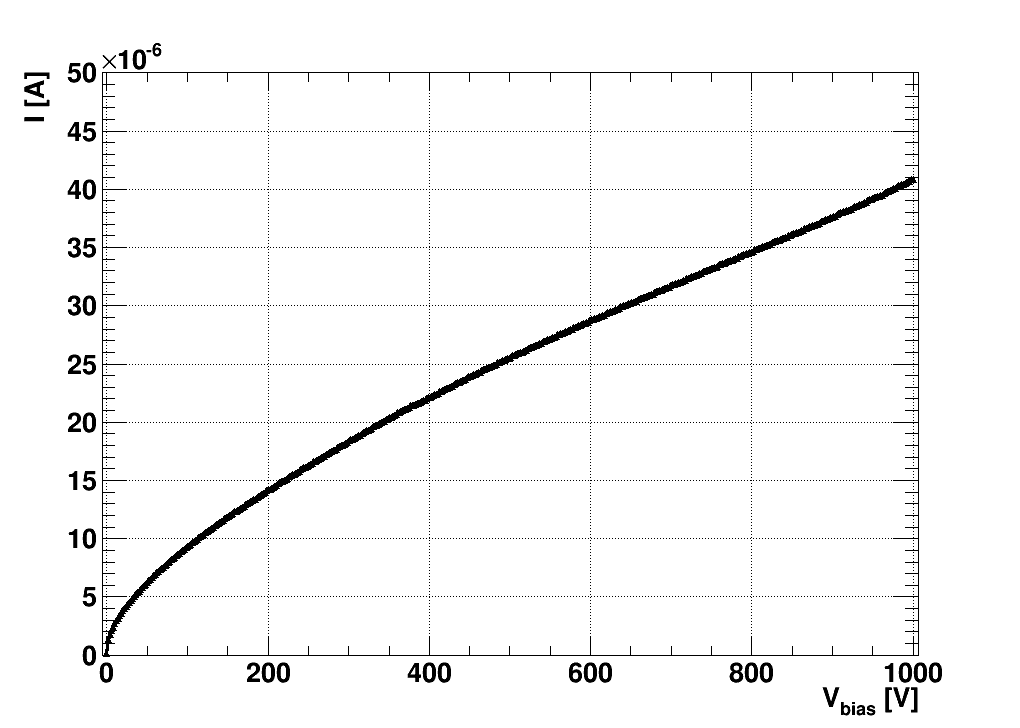}
% where an .eps filename suffix will be assumed under latex, 
% and a .pdf suffix will be assumed for pdflatex; or what has been declared
% via \DeclareGraphicsExtensions.
\caption{Leakage current as a function of the bias voltage  for the irradiated squared pad diode.  Both GRs and the pad were kept at ground potential while the backside was 
biased via a p$^+$ implant at the detector periphery. }
\label{fig:irrIV}
\end{figure}

The current voltage characteristic was studied for the same irradiated diode; the current flowing through the pad is reported in Figure~\ref{fig:irrIV}, while all the GRs were 
kept at the pad potential. 
After the irradiation the BD voltage value exceeds 1000~V; this is expected as the increased positive oxide charge depletes the p-spray, making then less pronounced the 
electric field peaks at the n$^{\rm +}$-p junctions. 
From the leakage current value at depletion voltage, the leakage current increase per unit of volume per unit of fluence  $\alpha$ is evaluated:
 $\alpha=\frac{I(\Phi)-I(0)}{V\Phi}$, where $I(\Phi)$ ($I(0)$) is the leakage current value after (before) the 
irradiation, $V$ is the sensor volume and $\Phi$ the total integrated fluence. A value of about $\alpha=4.7\times10^{-17}{\rm A/cm}$ is found which is in good agreement 
with the one reported in literature~\cite{moll-thesis} given the limited annealing of the samples after the irradiation.

\begin{figure}[!t]
\centering
\includegraphics[width=3.5in]{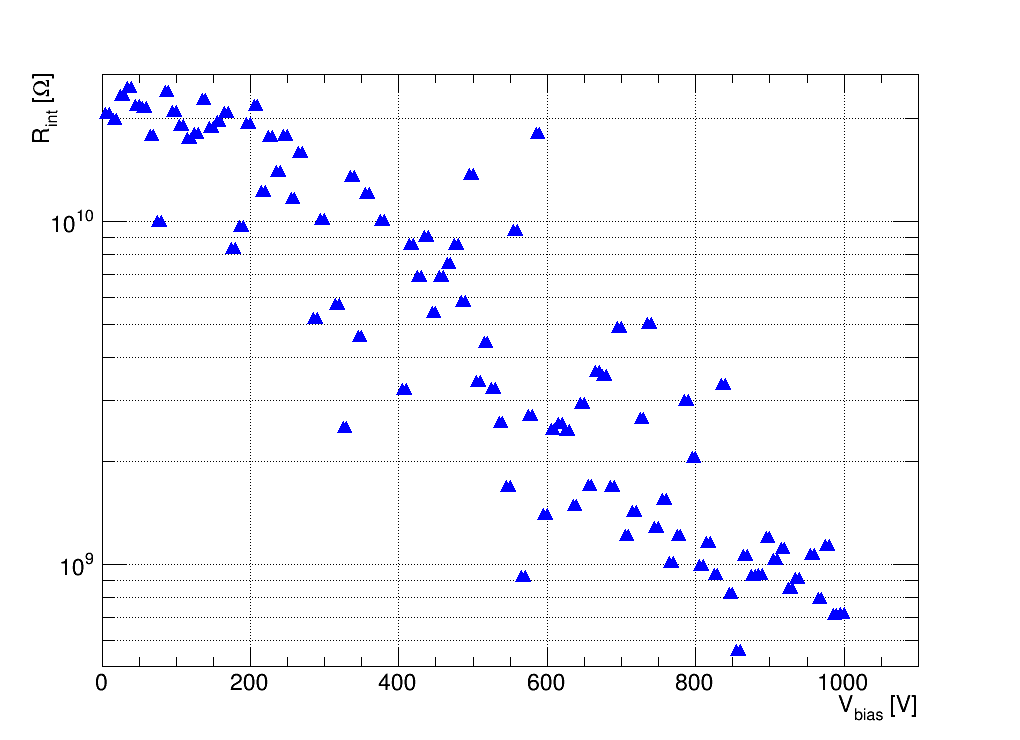}
% where an .eps filename suffix will be assumed under latex, 
% and a .pdf suffix will be assumed for pdflatex; or what has been declared
% via \DeclareGraphicsExtensions.
\caption{\label{fig:irrRint}Inter-pixel resistance as a function of the bias voltage.}
\end{figure}

To evaluate the degree of isolation among the pixels the effective inter-pixel resistance was evaluated on the inter-pixel structure. The substrate was polarized,
 in a series of voltage steps, at a negative voltage $V_B$ with respect to the central pixel; for each of these bias voltage steps, the pixels neighboring the central one had their
  voltage $V_N$ varied between -0.5~V and +0.5~V and the current $I_C$ flowing through the central pixel was registered; the central pixel was kept at ground all along the 
  measurement.  The inter-pixel effective resistance was  defined as: $R_{\rm int} = ({\rm d}I_C/{\rm d}V_N)^{-1}$. The measurement results are reported in Figure~\ref{fig:irrRint}. 
  As it can be seen the inter-pixel resistance $R_{\rm int}$ is in excess of 1~G$\Omega$ for almost all the considered bias voltage $V_B$ range (0-1000~V), which 
  indicates an high level of pixels isolation. This level of isolation after irradiation is not surprising as the samples were irradiated at a nuclear reactor where the 
  gamma flux is limited. There are plans to irradiate with charged particles more samples from this production and to perform again this study.

\begin{figure}[!t]
\centering
\includegraphics[width=3.4in]{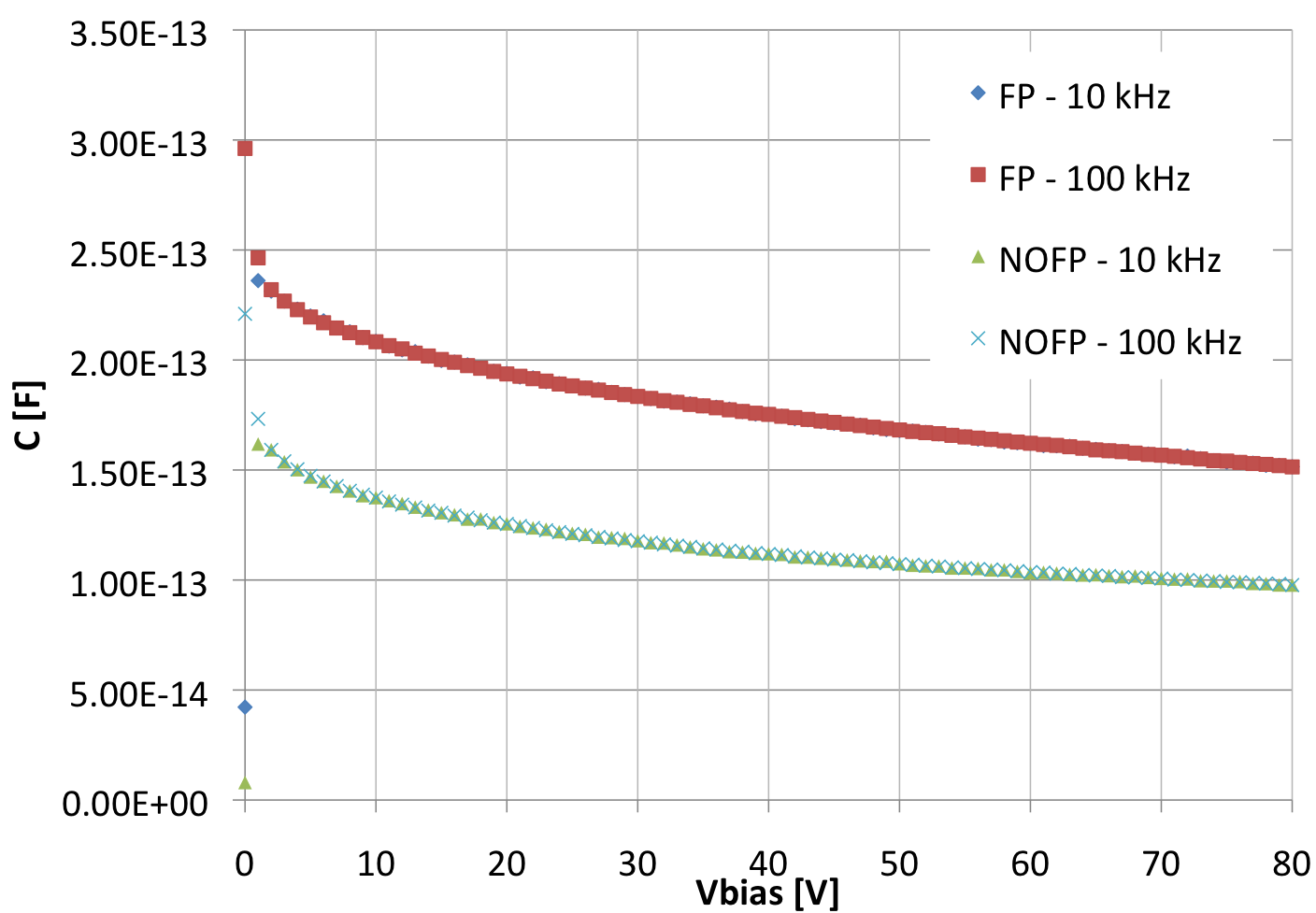}
\includegraphics[width=3.5in]{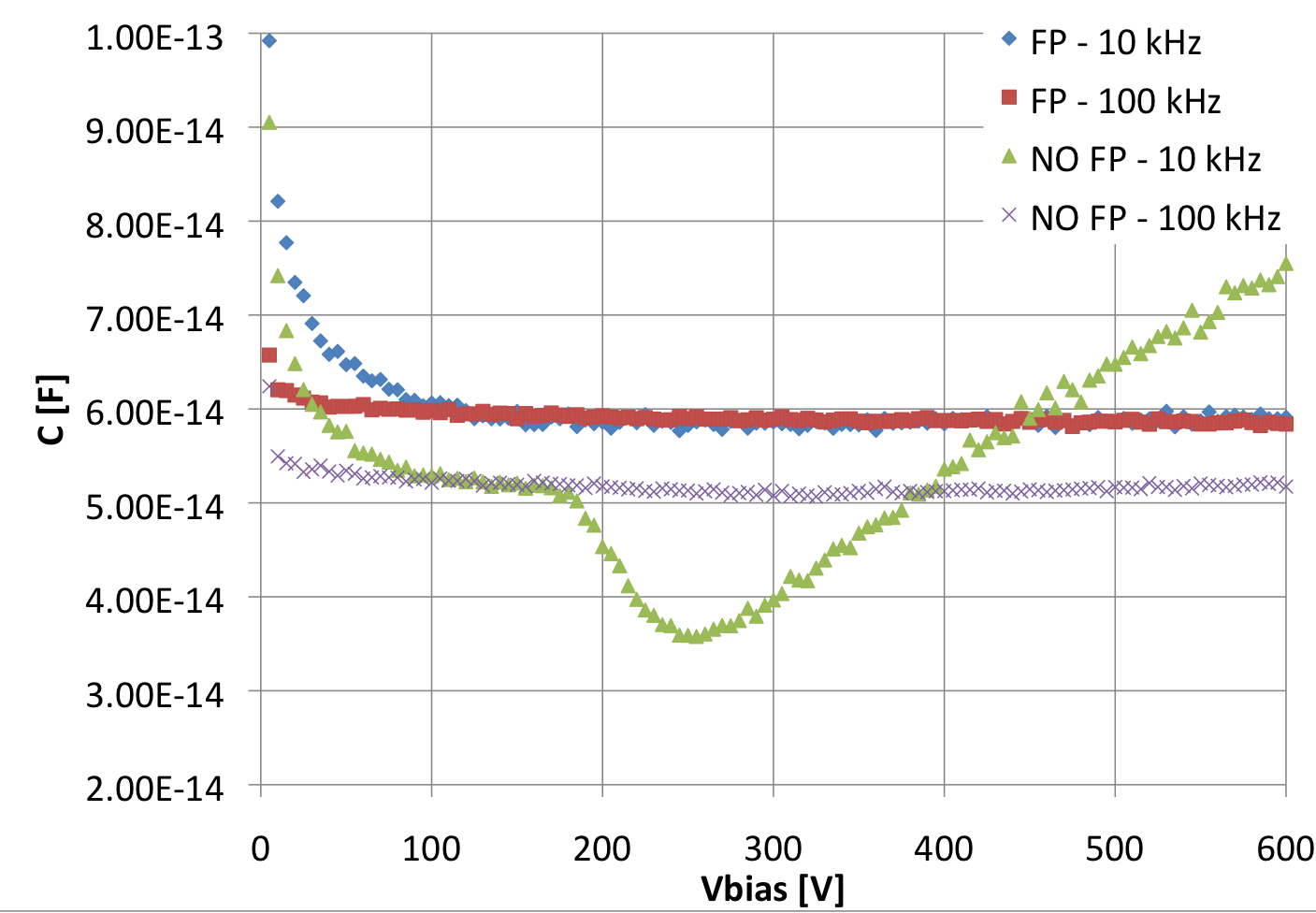}
% where an .eps filename suffix will be assumed under latex, 
% and a .pdf suffix will be assumed for pdflatex; or what has been declared
% via \DeclareGraphicsExtensions.
\caption{Inter-pixel capacitance as a function of the bias voltage, before (top) and after (bottom) irradiation; measurements were performed on pixels having or not a 
field-plate (FP), at different frequencies.}
\label{fig:irrIntCV}
\end{figure}

The inter-pixel capacitance was evaluated using the inter-pixel structure. Measurements were performed at two different frequencies (10 and 100~kHz) for pixels with and 
without the field-plate (see Figure~\ref{fig:pix_cap_struct} for details).  Measurement results are reported in Figure~\ref{fig:irrIntCV}, before (top) and after (bottom) 
irradiation.\footnote{The results for pixels after irradiation without field-plate measured at a frequency of 10~kHz (full triangles in Figure~\ref{fig:irrIntCV}, bottom), are reported for 
completeness, 
even if they are  meaningful  for bias voltages only up to 150~V.}
As expected after irradiation the inter-pixel capacitance is lower than before; this is due to the larger p-spray depletion due to the  radiation induced oxide charge density increase.
The presence of the field-plate translates into a capacitance increase of about 50~\% before the irradiation and of about~20~\% after it. 
A ten-fold increase in the frequency measurement seems to induce any change in the registered capacitance. Future measurements will  be carried  at larger 
frequency values (1~MHz and more). In summary, the level of capacitive coupling among neighboring pixels is at an acceptable level in term of capacitive input on 
the preamplifier in all the situations that were considered. 

\begin{figure}[!t]
\centering
\includegraphics[width=3.5in]{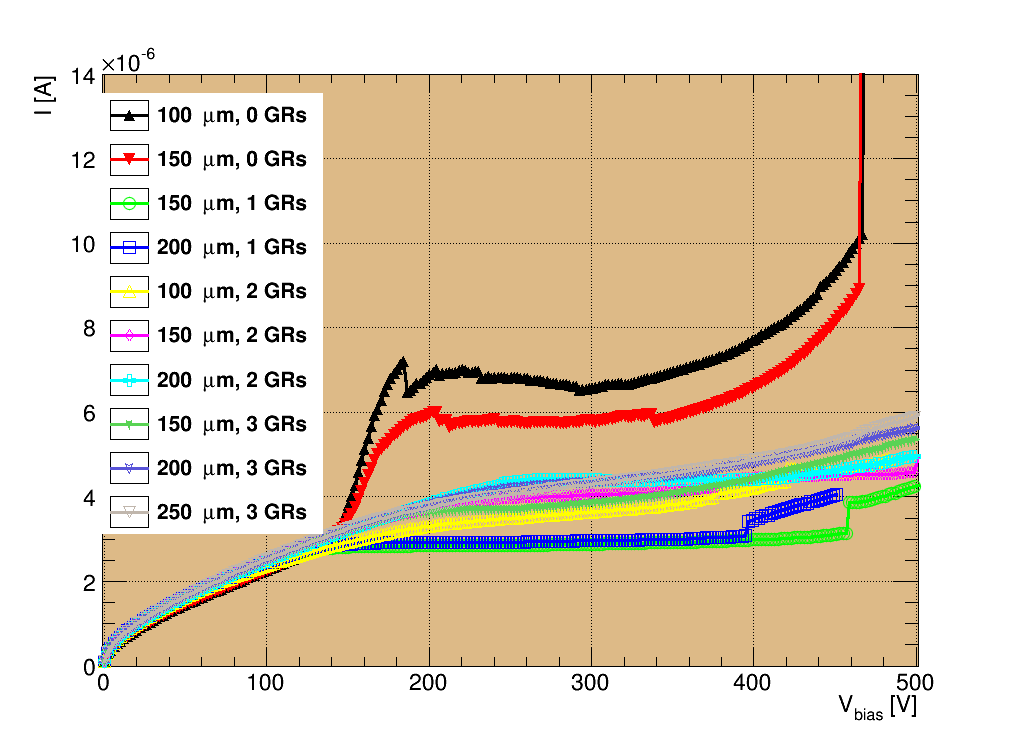}
% where an .eps filename suffix will be assumed under latex, 
% and a .pdf suffix will be assumed for pdflatex; or what has been declared
% via \DeclareGraphicsExtensions.
\caption{Pixels current as a function of the bias voltage for irradiated FE-I4 test structures; several sensors layouts are compared.}
\label{fig:irrIPad}
\end{figure}

\begin{figure}[!t]
\centering
\includegraphics[width=3.5in]{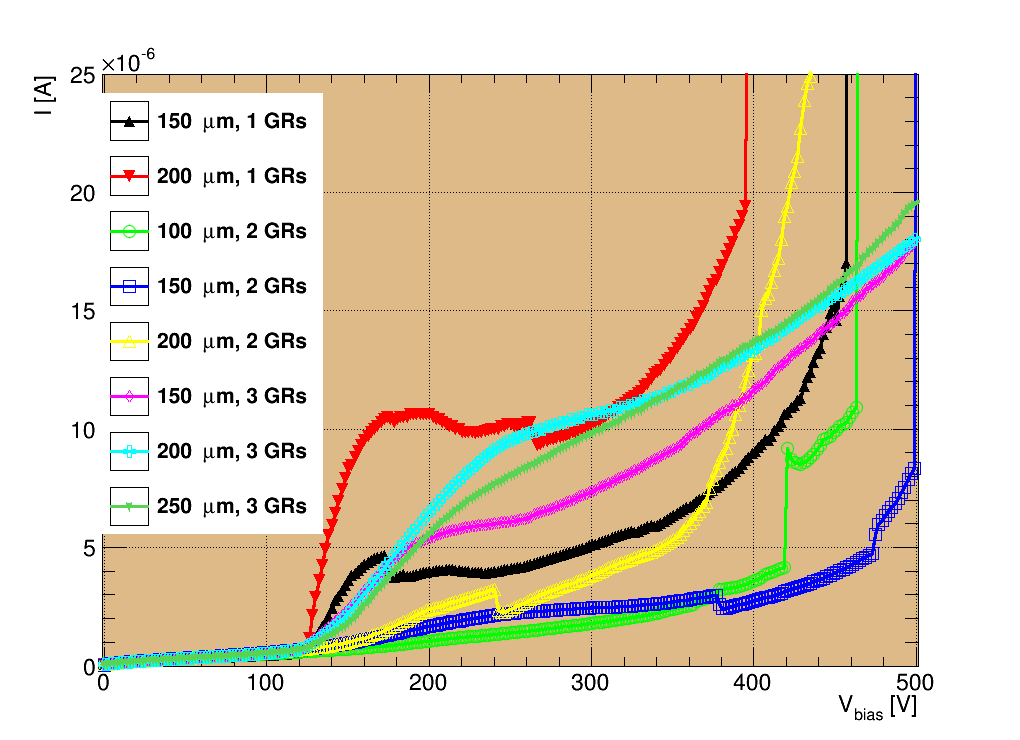}
% where an .eps filename suffix will be assumed under latex, 
% and a .pdf suffix will be assumed for pdflatex; or what has been declared
% via \DeclareGraphicsExtensions.
\caption{Innermost GR current as a function of the bias voltage for irradiated FE-I4 test structures; several sensors layouts are compared.}
\label{fig:irrIGR}
\end{figure}

The FE-I4 test structures were used to evaluate the BD voltage. The substrate was biased through the bias tab at a negative voltage with respect to the pixels and the 
innermost GR. The current flowing through the pixels and the innermost GR was registered. Several combinations of pixet-to-trench distance and number of GRs were tested; 
the complete list is reported in Table~\ref{tab:irrIvTable}.

\begin{table}[!ht]
\caption{\label{tab:irrIvTable}The different FE-I4 test structures under test after irradiation.}
\centering
\begin{tabular}{cc}
\# of GRs & pixel-to-trench distance (${\rm \mu m}$) \\
\hline
  0 & 50 \\
  0 & 100 \\
  0 & 150 \\
 1 & 100 \\ 
 1 & 150 \\ 
 2 &100\\
 2 & 150 \\ 
 2 & 200 \\ 
 3 & 150\\ 
 3 & 200\\ 
 3 & 250\\ 
\end{tabular}
\end{table} 

Sensors featuring 50~${\rm \mu m}$ as  pixet-to-trench distance and no GRs and 100~${\rm \mu m}$ as  pixet-to-trench distance and one GR weren't working properly after 
irradiation so they won't be included in the results.
In Figure~\ref{fig:irrIPad} the current flowing through the pixels is reported. It can be seen that sensors having no GRs show a BD voltage value before the depletion voltage; 
anyway, those sensors can be operated at bias voltages up to 400~V which should be enough to detect a signal from a minimum ionizing particle (MIP)\footnote{This has to be 
confirmed in a foreseen beam test.}. For sensors having at least one GR the pixels current is very stable within 300 and 500~V; this is very promising for real FE-I4 sensors 
in view of the HL-LHC.

In Figure~\ref{fig:irrIPad} the current flowing through the innermost GR is reported. It can be seen that adding more GRs increases the BD voltage as expected. The result of the 
sensor featuring 150~${\rm \mu m}$ as as  pixet-to-trench distance and one GR is quite remarkable: it can be operated at a bias voltage up to 500~V with a seven-fold 
reduction of the edge's dead area width with respect current ATLAS pixels~\cite{AtlasPixels}. 
Sensors with 3 GRs show no BD up to 500~V which is another excellent result for sensors having edge's dead area width as low as 150~${\rm \mu m}$.

\section{FE-I4 pixels module assembly}\label{sec:fei4}

FE-I4 sensors are being bump-bonded at IZM\footnote{Fraunhofer-Institut f\"ur Zuverl\"assigkeit und Mikrointegration - Berlin, Germany} to FE-I4B read out chips. 

\begin{figure}[!t]
\centering
\includegraphics[width=3.5in]{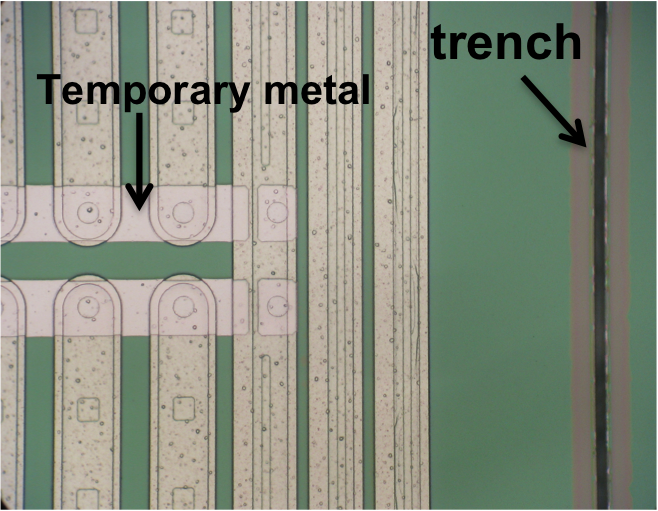}
\includegraphics[width=3.5in]{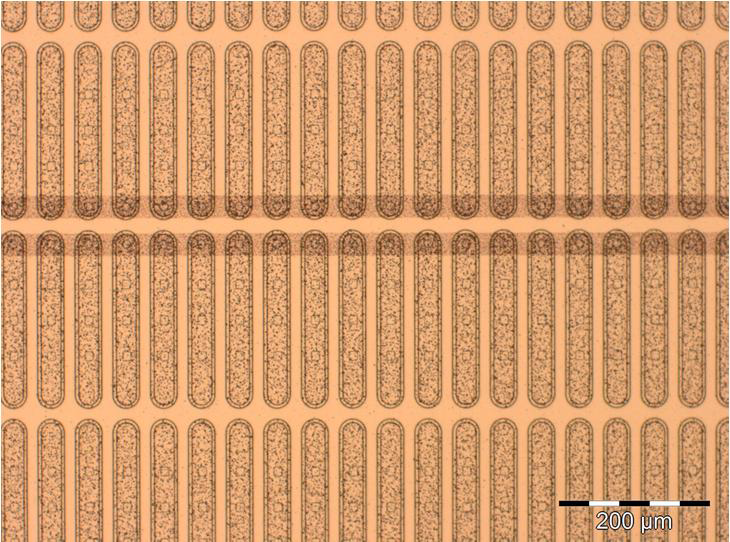}
% where an .eps filename suffix will be assumed under latex, 
% and a .pdf suffix will be assumed for pdflatex; or what has been declared
% via \DeclareGraphicsExtensions.
\caption{Pictures of the pixels-side of a FE-I4 sensor. Top: sensors before temporary metal removal; region covered with temporary metal and the trench are highlighted. Bottom: 
Same sensor as above (different scale, though)  after temporary metal removal.}
\label{fig:tempmetal}
\end{figure}

It is necessary to select good FE-I4 sensors at the wafer level for bump-bonding, by measuring their I-V characteristics.
 For this purpose, an additional layer of metal~\cite{6522814} was deposited over the passivation and patterned into stripes, each of them shorting together a 
 row of pixels, contacted through 
 the small passivation openings foreseen for the bump bonding.
After the automatic current-voltage  measurement 
 on each FE-I4 sensor (reported in~\cite{2013arXiv1311.1628B}), the metal has been removed by  wet etching, which does not affect the electrical characteristics  of the devices.
Pictures of FE-I4 sensor pixels before and after the metal layer removal can be seen in Figure~\ref{fig:tempmetal}. 

\begin{figure}[!t]
\centering
\includegraphics[width=3.5in]{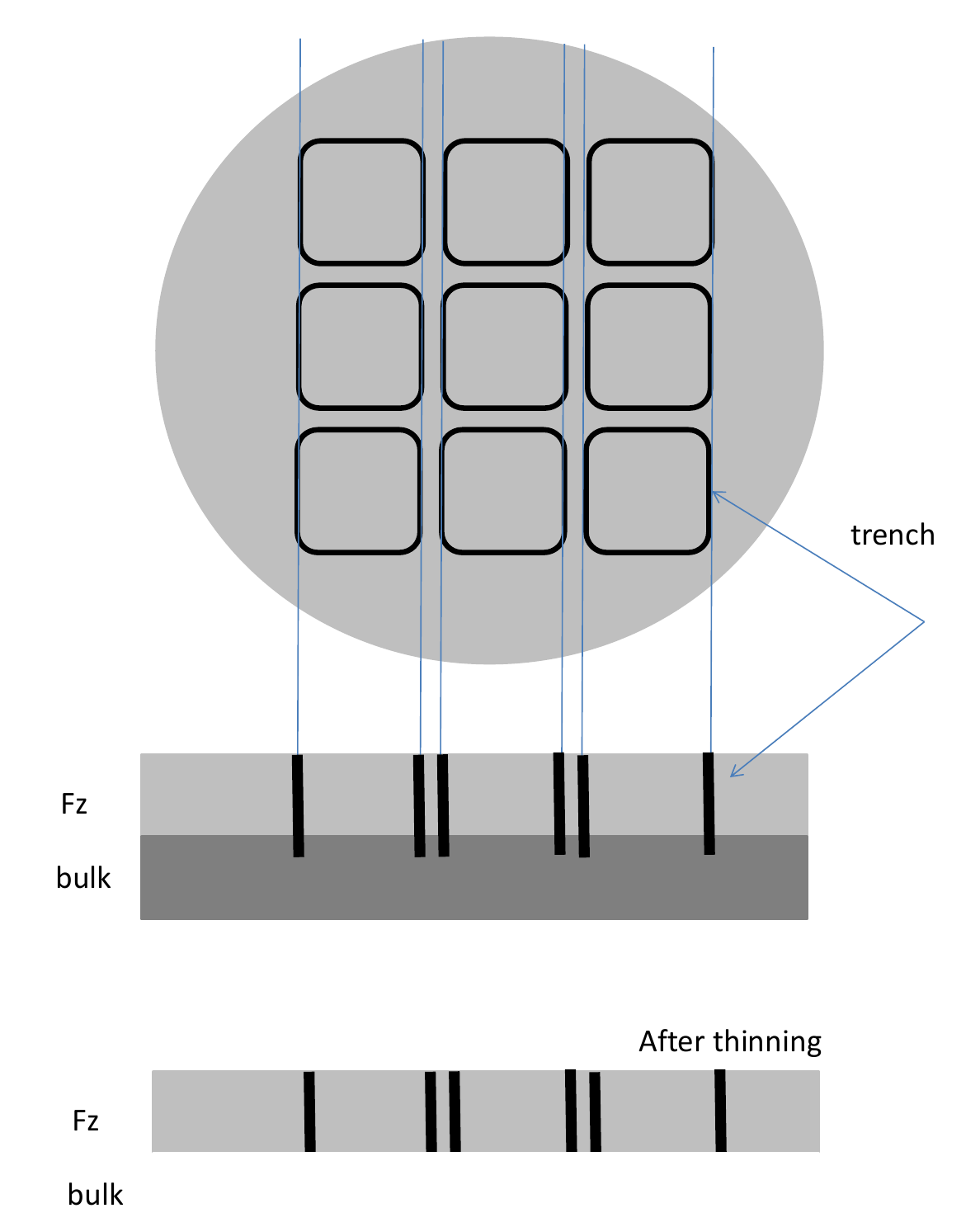}
% where an .eps filename suffix will be assumed under latex, 
% and a .pdf suffix will be assumed for pdflatex; or what has been declared
% via \DeclareGraphicsExtensions.
\caption{Sensors separation from wafer.}
\label{fig:lapping}
\end{figure}

To further proceed in module construction  the support wafer has to be removed. The approach followed is illustrated in Figure~\ref{fig:lapping}. Each FE-I4 sensor is 
surrounded by the trench on all sides, so  the sensor is effectively isolated  on all sides from the silicon wafer. 
After having deposited a dicing tape on the pixel side, the support wafer can be back-lapped completely. Since the trench penetrates  the whole sensor wafer thickness, 
once the support wafer has been completely lapped, each sensor can separated from the others by removing the tape.

The process will then follow the usual steps for hybrid pixels module assembly.

\section{Conclusions and outlook}\label{sec:concl}

In view of the upgrade of the ATLAS Inner Detector for HL-LHC runs,
 FBK Trento and LPNHE Paris developed new planar n-on-p pixel sensors, characterized by a reduced inactive region at 
 the edge thanks to a vertical doped lateral surface at the device boundary, the ``active edge'' technology. 
 The  measurements performed on irradiated samples, including capacitance- and current-voltage characteristics, show that it is possible to operate them successfully   
after irradiation. A good level of pixels isolation is reached even at low p-spray doses after irradiation. 
Functional tests of the pixel sensors  with radioactive sources, before and after irradiation, 
 and eventually in a beam test, after having bump bonded a number of  pixel sensors to the FE-I4 read out chips, will follow.

\section*{Acknowledgments}

We acknowledge the support from the MEMS2 joint project of the Istituto Nazionale di Fisica Nucleare and Fondazione Bruno Kessler.

%\bibliographystyle{IEEEtran}
%\bibliography{IEEEabrv,biblio.bib}


\end{document}